\documentclass[prd,aps,floats,twocolumn]{revtex4}

\newcommand{\be}{\begin{equation}}
\newcommand{\ee}{\end{equation}}
\newcommand{\ba}{\begin{eqnarray}}

\newcommand{\ea}{\end{eqnarray}}

\newcommand{\bc}{}

\newcommand{\bra}[1]{\left(#1\right)}

\usepackage{mathrsfs}

\usepackage{amssymb}

\usepackage[dvips]{graphicx}

\usepackage{wrapfig}

\begin{document}

\preprint{\
\begin{tabular}{rr}
\end{tabular}
}

\title{Conservative Estimates of the Mass of the Neutrino from 
Cosmology}

\author{C.~Zunckel and P.G~Ferreira}
\affiliation{Astrophysics Physics, University of Oxford, Denys Wilkinson 
Building, Keble Road, Oxford OX1 3RH, UK}
\begin{abstract}
A range of experimental results point 
to the existence of a massive neutrino. The recent high precision 
measurements of the cosmic microwave background and the large scale surveys of 
galaxies can be used to place an upper bound on this mass. In 
this paper we perform a thorough analysis of all assumptions that go into obtaining a 
credible limit on $\sum m_{\nu}$. 
In particular we explore the impact of extending parameter space beyond the current standard cosmological model, 
the importance of priors and the uncertainties due to biasing in large scale structure.   We find that the mass constraints are independent of the choice of 
parameterization as well as the inclusion of spatial curvature.   The results of including the possibility of dark energy and 
tensors perturbations are shown to depend critically on the data sets used.
The difference between an upper bound of $2.2$ eV, assuming generic initial conditions, 
and an upper bound of $0.63$ eV, assuming adiabaticity and a galaxy bias of $1$, 
demonstrate the dependence of such a constraint on the assumptions in the analysis. 
\end{abstract}
\date{\today}
\pacs{PACS Numbers : }
\maketitle

\noindent

\section{Introduction}
The neutrino is an integral component of the standard model of
particle physics. Until recently it was assumed to be
massless.  With the new advances in
non-accelerator particle physics there is now definitive evidence
that this cannot be so: neutrinos must have mass. The first signatures for their 
masses were observed flavour oscillations in atmospheric and solar neutrinos and have later been 
verified in accelerator and nuclear reactor sources \cite{KAM_all} . When these observations are interpreted within the 3-neutrino type scenario of the standard model, 
their masses are required to be of the order of the measured mass differences, leading to two possible mass scales; 
$\delta m_{atm} \simeq 3 \times 10^{-3} $ eV$^2$ and , in conjunction with the 
nuclear reactor experiment results from KamLAND, $\delta m_{sol} \simeq 5 \times 10^{-5} $ eV$^2$.  These mass scales can be accommodated by a
model in which at least two eigenstates have mass. Alternatively if their true mass exceed $0.1$eV then these results indicate that all three species are nearly degenerate. 
Tritium decay measurements have also been able to place an upper limit on the
electron neutrino mass of $2.3$ eV at the 95$\%$ confidence
level (CL) \cite{Bonn}.  Although difficult to reconcile with the rest of the data, it is interesting to note 
the result from the Los Alamos Liquid Scintillator Neutrino Detector (LSND) 
which implies an lower limit of $m_{\nu}>0.4$ eV \cite{LSND}. 
Given our certainty that they exist, these results make the neutrino one of the most
convincing dark matter candidates \cite{Lesgourges}.

Although it is unlikely that these massive neutrinos are the dark matter it 
appears conceivable that they may
still affect the growth of density perturbations
in a measurable way.  Neutrinos with a mass less than $2$ eV are still relativistic 
when entering the horizon for scales of
$k=0.1$ h Mpc$^{-1}$ and are quasi-relativistic at recombination. 
Therefore they cannot be treated as a non-relativistic component of the CMB 
and are not entirely degenerate with the other relativistic components 
\cite{seljak}. In a matter dominated universe, a modicum
of massive neutrinos will free-stream on scales of clusters and 
galaxies and therefore suppress the rate of growth  of 
density perturbations from being proportional to the scale factor 
$a$ to being proportional to $a^{1-\epsilon}$ where 
$\epsilon=\frac{5}{4}[1-(1-24\Omega_\nu/25)^{1/2}]$, 
where the neutrino mass, $\sum m_\nu$ is related through the expression:
$\Omega_\nu=\sum m_\nu/(93.15 h^2)$. 
Here the Hubble constant today is $H_0=100h$km s$^{-1}$ Mpc$^{-1}$.
This relation can be applied even in the event that the neutrino
species are found to be non-degenerate \cite{Lesgourges}.
This effect on the growth of structure supplies us with a useful 
method for constraining $\Omega_\nu$ and as a result, $\sum m_\nu$. By
measuring the amplitude of clustering on large scales (above the
free streaming scale) and comparing it to the level of clustering
on small scales (below the free streaming scale) it is possible
to tease out the level of damping due to the neutrinos. The
amount of clustering on large scales is well constrained by
measurements of fluctuations in the Cosmic Microwave Background
(CMB) which map out the density perturbations on scales up to
the horizon.  Surveys of galaxies allow us to
pin down the amount of clustering on small scales. Combining the two 
allows us to place a constraint on $\Omega_\nu$.

This approach has been applied extensively over the last decade.
 In \cite{WMAP3_spergel},  
using a combination of data from the Wilkinson Microwave Anisotropy 
Probe 3-year (WMAP3) \cite{WMAP3}, 2dF Galaxy Redshift Survey \cite{2dF_05}, 
Sloan Digital Sky Survey (SDSS) \cite{SDSS} and the supernova (SN) \cite{riess, astier}, 
the WMAP team placed an upper limit of
$\sum m_{\nu}<0.66$ eV on the neutrino mass when
the SDSS bias constraint \cite{seljak} is included. 
Seljak \cite{seljak_06} recently reported an huge improvement on this when the same dataset 
is supplemented with the
baryon acoustic oscillation (BAO) \cite{BAO} and Lyman-$\alpha$ \cite{ly} constraints.
The Lyman-$\alpha$ forest provides information about the matter power spectrum on small scales 
(where neutrinos suppress power) and 
at high redshift ($z=2 - 4$) where the non-linear evolution is less significant.  
The Lyman-$\alpha$ data prefers a higher normalization and
is the primary source of the strong upper bound on the neutrino mass of $\sum m_{\nu} < 0.17$ eV at 95$\%$ CL 
in the absence of the SDSS bias constraint.  This constraint is found using the 
recent measurement of the position of a peak in the galaxy 
correlation function at $\sim 140$ Mpc h$^{-1}$ arising from the baryon
acoustic oscillations once they have decoupled from photons.
Given its dependence on $\Omega_m$ and the hubble parameter $H(z)$ 
and the degeneracy of neutrino mass with both these parameters the use of the SN data to fix $\Omega_m$ in conjunction with the information 
about $\Omega_{\Lambda}$ provided from the BAO result proves to be powerful.

The use of cosmological observations can, in principle, supply us with
limits on the neutrino mass which are comparative with 
experimental bounds. Yet there is a valid concern that
cosmological constraints depend largely on the chosen cosmological model and accompanying assumptions/priors . The
impact of certain assumptions that go into such an analysis has been considered by other authors 
\cite{Hannestad2, EL2}, however a detailed review of 
all relevant factors that may be getting in the way of 
a truly robust constraint. In this paper we wish to elucidate how robust the neutrino constraints are to the assumed model. 
The structure of this letter is as follows. We first establish the details of our
the analysis and determine the way in which choices of prior probability distributions for the neutrino 
content affect our results within the context of a $\Lambda$CDM universe. We then extend the model to include the 
possibilites of dark energy, curvature, tensors, spectral running and more general initial conditions.  We then consider 
how knowledge of cosmological parameters such as the galaxy bias can be harnessed to improve the neutrino mass limits.

\section{The approach}

Neutrinos will impact cosmological observables in a variety of ways.
Relativistic neutrinos have a marginal effect by increasing the radiation density before decoupling thereby impinging on the acoustic peak morphology.  
Massive neutrinos that become non-relativistic prior to recombination suppress the early integrated Sachs-wolf effect
Their signature is a modification of the height and position of the first peak.
Furthermore, the free-streaming of massive neutrinos induces a more rapid decay of the gravitational potential, 
fueling the acoustic oscillations within this free-streaming scale. This modifies the heights of 
the second and third peaks relative to the first which can not be reconciled in the context of the shift in peak position and facilitates the constraint. 
As shown in \cite{Ichikawa} it is thus possible to derive limits on neutrino mass from the CMB alone, however
these above mentioned degeneracies weaken the power of the data and the constraints come out to be $\sum m_{\nu} < 2.0$ eV for flat power-law $\Lambda$CDM. 
The merits of this are that the limit suffers less from systematic errors and is robust
since it is derived from the single experiment with one set of systematics. The most effective use of the CMB data in pining down $\sum m_{\nu}$ is however 
in the normalization of the large scale galaxy clustering power. The primary effect of the neutrino mass on the matter power spectrum $P(k)$ is to 
reduce power on scales smaller than the free-streaming scale. 
Using both LSS and CMB data concurrently has been identified as the means of arriving at reliable and competative constraints. 

The empirical relationship between the morphology of the supernova type 1a light curves and their intrinsic 
luminosities facilitates their use as standard candles. These datasets \cite{riess, astier} have become key in determining the expansion rate of the universe and thus in 
testing models of dark energy. However the physical mechanism that forms the basis for this relation has yet to be established and given the enormous impact 
of these measurements we choose not to include them in our standard data set but will incorporate the SN results taken from \cite{astier} at points for the sake of comparison.

We approach the Lyman-$\alpha$ measurements  \cite{ly} with even more caution given the as yet still 
unresolved systematics that still plague these data sets.  When re-analysed by \cite{seljak_06} and 
\cite{Viel} discrepancies arose, with the authors of \cite{Viel} finding a lower normalization than the primary result in \cite{ly}.   Since the 
Lyman-$\alpha$ data has been shown to  tighten the bound on neutrino mass considerably we choose not to include it in our analysis. 
The impact of the inclusion of these different datasets on the neutrino mass 
limits has been comprehensively reviewed in \cite{Kristiansen} and found to be significant.

The practicalities of the method are as follows. CMB polarization and temperature power 
spectra as well as the matter power spectra are computed using the CAMB package 
\cite{LCL}.  We compare the spectra computed from the sample model to our fiducial dataset, comprising 
of the CMB temperature anisotropy measurements from WMAP 3-year \cite{WMAP3}
and a combination of small scale (high $\ell$) CMB data from ACBAR, BOOMERANG, CBI and VSA \cite{small}, 
complemented with the galaxy power spectrum derived from the 
SDSS \cite{SDSS}. We additionally use the constraint on the baryon density today of $\Omega_{b} h^2$
from Big Bang neucleosynthesis (BBN) \cite{Burles, Freedman}. In \cite{Steigman} tension between the 
WMAP 3-year data and the BBN measurement was identified, however this prior is necessary to exclude 
wild high $\Omega_{b} h^2$ cosmological models which are favoured when isocurvature is allowed. 

A likelihood analysis using the likelihood function in
\cite{Verde-et al} is then performed in order to compare the 
spectra generated from the models with the data. 
We use a Monte Carlo Markov chain method that invokes a 
Metropolis algorithm as described in \cite{dunkley, dunkley2}
to explore the resulting likelihood distribution in
parameter space efficiently. 

We take the concordance model as our starting point: a spatially 
flat Universe with nearly scale invariant adiabatic fluctuations 
dominated by
cosmological constant.  A Fisher matrix analysis \cite{Hannestad2} reveals strong degeneracies 
between the neutrino density $\Omega_{\nu}$, the galaxy bias $b$ 
and the matter density
$\Omega_m=\Omega_b +\Omega_d$. These parameters impact matter power spectrum strongly on scales smaller
than the free-streaming scale, while their effects on the CMB are similar in their subtlety (in the case of the bias, none at all).  
Neutrinos with masses of order eV become non-relativitistic only after decoupling, when the evolution 
determining the shape of the CMB spectrum has already taken place \cite{Lesgourges}.

The bias parameter shares an effect on the normalization of the power spectrum on all scales so an increase in $\sum m_{\nu}$
could be partly compensated for by a decrease in $b$. 
Increasing the matter content $\Omega_m$ brings
matter domination forward in time, resulting in less suppression
on small scales ( i.e. shallower gradient at larger $k$ \cite{HET}) and 
a smaller horizon at matter-radiation equality \cite{Hannestad}.  It then 
makes sense that the power suppression at larger $k$ 
caused by the massive neutrinos must be corrected for this opposing effect; 
$\frac{\Delta P}{P}\simeq -8\frac{\Omega_{\nu}}{\Omega_{m}}$. 
A change in $\Omega_m$ is equivalent 
to a change in $\Omega_{\Lambda}$ and the
degeneracy between this parameter and the neutrino mass. 

Adding in 3 types of massive neutrinos degenerate in mass (we in fact use $N = 3.04$  
as predicted by theory \cite{seljak}), 
the cosmological parameters that sufficiently describe this scenario 
are the physical baryon density $\omega_b$, the physical 
cold dark matter (CDM) 
density $\omega_d$, the neutrino density relative 
to CDM $f_{\nu}=\omega_{\nu}/\omega_{d}$,
the fractional density of cosmological constant $\Omega_{\Lambda}$, 
the galaxy bias $b$ (assumed to be a constant based on theoretical reasoning in \cite{Benson}), 
the amplitude and scalar spectral index of the primordial 
fluctuation spectrum $A_s$ and $n_s$ respectively. 

The parameterization used by the WMAP team $f_{\nu}$ is the 
ratio of the density of this hot dark matter component to the CDM.  Given the degeneracy between $\omega_{\nu}$ and $\Omega_d$ 
the ratio of the two is likely to be a sensitive parameter.
There are however alternatives; in other work the parameterization is chosen to be $F_{\nu}$, the fraction of the 
total matter density $\Omega_M=\Omega_b+ \Omega_d+\Omega_{\nu}$ that the neutrinos 
comprise. In other cases one simply works with the total mass, $\sum m_{\nu}$, of  all 3 species.  
It seems worthwhile to check the effect of the neutrino parameterization by
repeating the same analysis using $\sum m_{\nu}$ 
and $F_{\nu}=\frac{\Omega_{\nu}}{\Omega_M}$ as the variables. 
If seemingly benign flat priors are imposed different parameterizations the 
effective priors on the quantity being constrained, in this case $\sum m_{\nu}$, will diverge. 
Taking the priors shown in table \ref{param_table} which are typically placed on the various neutrino variables in such work, 
we find that the upper limits on $\sum m_{\nu}$ at the 95$\%$ CL (using the CMB and LSS data alone) are in agreement for all 3 parameterizations.  
In the region of $\sum m_{\nu}$ space in which the data falls ($<4$ eV), 
these priors are all effectively equivalent top-hat and thus have no impact.  
If we add the SN data \cite{astier} which amounts to imposing the constraint $\Omega_m=0.263 \pm 0.042$, 
the effective posterior probabilities look quite different in this region of $\sum m_{\nu}$ space, but again do not lead to major 
discrepencies in the mass constraints. The situation does not change 
when $\sigma (\Omega_m)$ is reduced to $0.01$ as predicted for future experiments such as SNAP.
This is reasonable given that $\Omega_M$ and $\Omega_b$ are well constrained by the current data. 
We conclude the use of different subsets of the current data does not affect the 
independence of parameterization on the neutrino mass limits and thus continue using $f_{\nu}$ throughout the rest of the analysis.

We now explore deviations from the standard model, specifically 
spatial curvature characterized by $\Omega_K$,  the presence of dark energy, 
the presence of tensors, the running of the spectral index and isocurvature contributions to the initial fluctuation spectrum. 

\begin{table}
\begin{center}
\begin{tabular}{ccc}    \hline
\textbf{Parameterization}  & \textbf{Prior} & $\sum m_{\nu}$ for $N_{\nu}=3.04$ \\
 & & $95 \%$ CL 
\\ \hline
$ f_{\nu} = \Omega_{\nu}/  \Omega_d$ &  $0 - 1$ & $1.26$ eV\\
 $F_{\nu} = \Omega_{\nu}/  \Omega_M$ &  $0 - 1$ & $1.27$ eV\\
$\sum m_{\nu}$ &  $0 - 5$ eV & $1.29$ eV \\
\hline
\end{tabular}
\end{center}
\caption{Constraints on neutrino mass when different parameterizations are used.}
\label{param_table}
\end{table}

\section{Extending Parameter space beyond $\Lambda$CDM}
Despite the vast amount of data now in support of the simple
concordance picture, more complicated scenarios are possible in
certain paradigms. In order to be certain that the least restrictive
mass limit is obtained, the model-dependence of these constraints should be explored. 

\subsection{Dark energy?}
\begin{figure}[t]
  \begin{center}
    \includegraphics[trim = 0mm 0mm 0mm 0mm, scale=0.7]{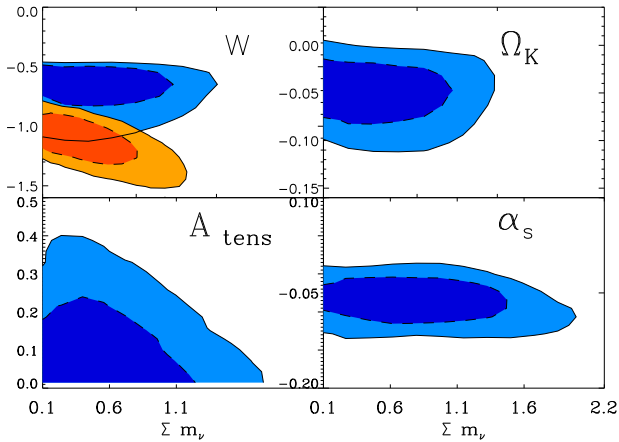}
    \caption{Extending the model beyond $\Lambda$CDM.  \emph{From top right, bottom left to right:}  95 and 68 $\%$ contours in the ($\sum m_{\nu}$, $\Omega_K$), 
    ($\sum m_{\nu}$, $\alpha_{s}$) and  ($\sum m_{\nu}$, $r$)  planes.  \emph{Top left:} The confidence intervals in the ($\sum m_{\nu}$, $w$) 
    planes for our CMB+LSS dataset (blue) and for CMB+LSS+SN dataset (red). }
    \label{test4}
  \end{center}
    \vspace{.1in}
\end{figure}

In recent work \cite{ WMAP3_spergel, hannestad_w, goobar} 
a degeneracy between the equation of state of 
dark energy $w$ and the neutrino mass was identified. Massive neutrinos contribute to the energy 
density of the matter component in the universe and thus can potentially alter the epoch of matter-radiation 
equality as well as the diameter angular distance to the last scattering surface. We wish to
clarify this result.  If $w$ is allowed to vary, the effect of larger neutrino masses can be 
mimicked in the power spectrum by more negative values of $w$ 
instead of larger $\Omega_m$ values which become incompatible with the SN data \cite{hannestad_w}.  If we make no assumption about the cosmological
constant and include a constant $w$ in the parameter space, the 
resulting confidence contours for the CMB and LSS dataset (disregard the SN results)
show no signs of a relationship between these parameters.  
The inclusion of the SN data \cite{astier} tightens the correlation between w and $\Omega_{\Lambda}$ and 
hence links more strongly to $\sum m_{\nu}$, giving rise to the degeneracy between that has been pointed out.  
This contrast is demonstrated in figure (\ref{test4}).  The investigation of the correlation between neutrino mass and 
dark energy has been extended in \cite{Xia} to include dynamical models in which the equation of state is characterized by a smooth time-varying function. 
Dynamical dark energy will impact the time-evolution of the gravitational potential wells and thus contribute to late time integrated Sachs Wolf (ISW) 
effect which will in turn manifest in the CMB power spectrum. A strong anti-correlation between $\sum m_{\nu}$ and the time-dependent part of 
$w(z)$ was found which accounts for the substantial weakening of  the upper bound to $\sum m_{\nu} < 2.8 $ (depending on inclusion/exclusion of dark 
energy perturbations) for the CMB+LSS+SN dataset.  As our understanding of dark energy improves, we may have more reason to use more complex 
models of this energy density component into neutrino mass analyses.

\subsection{Spatial Curvature?}
The analysis so far has been performed under the assumption of flat space.
We now consider the possibility of non-flat geometry and include the 
curvature density $\Omega_K$ as a parameter.
When we admit spatial curvature, a slightly closed universe is 
favourable with a best-fit model of $\Omega_K=-0.05$. A tighter 
constraint on the neutrino mass of $1.17$eV is also obtained.
The CMB feels the impact as the inclusion of $\Omega_{K}$ which dictates the position of the first acoustic peak. 
Since $\Omega_m h^2$ also affects all the peak positions, the favouring of an open Universe by the data leads to a higher 
distribution in $\Omega_m h^2$. But because of the increased freedom on the right-hand side of $\Omega_m + \Omega_r=1-\Omega_K$ which holds for
early times, a higher $\Omega_m$ is countered by a more negetive $\Omega_K$, producing a more gentle rise in the 
matter-radiation ratio and thus disfavouring the high $\sum m_{\nu}$ models.
The results are however consistent with flat space 
and massless neutrinos within 2-$\sigma$.  The confidence contour in
figure (\ref{test4}) rules out any degeneracy between the neutrino mass and 
$\Omega_K$. This parameter does not have notable effects on the allowable models and can be confidently ignored from the parameter space. 

\subsection{Running spectral index?}
Inflationary cosmological scenarios seem to provide the most 
convincing explanations for the initial conditions for structure formation \cite{chung_05}. 
Although a scale-invariant spectrum
of primordial fluctuations is regarded as an intrinsic
accompaniment of these models, this picture is only
valid within certain cosmologies, such as pure de Sitte
\cite{chung}. There have been claims of strong observational indications from the CMB, LSS and Ly-$\alpha$ data that the spectrum is a 
function of $k$ (e.g. see \cite{peiris}), however the fit of combinations of other datasets have been shown to deteriorate when running is 
allowed \cite{seljak}.  Introducing running also means that the constraints on other parameters are weakened and that the contribution 
of gravity waves can no longer be probed using the CMB data at all.
In particular, there exists a strong degeneracy between $\Omega_m$ and variations in the spectral slope on the scales probed by the CMB \cite{Lesgourgues, Ichikawa},\cite{Fukugita}.
A spectral index which decreases on small scales (negative running) would limit the amount of small-scale damping that could be 
tolerated from massive neutrinos. Given that significant running of the spectral index is still under debate and has implications on 
neutrino mass constraints, it seems worthwhile to explore.  In recent work \cite{Feng} the neutrino mass constraints were found to be lowered 
in the presence of a scale-variant $n$ when the CMB, LSS and SN data 
was used. The SN data prefers a lower value of $\Omega_m$ which means that the effect of increased $\sum m_{\nu}$ 
can not be absorbed by an slight increase in the matter density. 
We consider a running spectral index in the presence of massive neutrinos 
\emph{without} the admission of tensors using our fiducial dataset which does not include the SN results. 
 In this analysis we assume the form of the power spectrum to be
\be
1+ \frac{d~\ln \Delta^2_R (k)}{d~\ln k} = n(k_0) + \frac{d~n_s}{d~\ln (k)} \ln (k/k_0) .
\ee and parameterize the running of the spectral using  $\alpha_s = d n_s / d \ln k $. The pivot scale is $k=0.05$ Mpc$^{-1}$. We find 
the constraint on $\sum m_{\nu}$ opens up to $1.66$ eV ($95\%$ C.L.) while the amount of running allowed comes 
out as $\alpha_s= -0.062 \pm0.02$ in agreement with the findings of the WMAP team using the 3-year data. 
This result, although at odds with the findings in \cite{Feng},
is expected given the signficantly higher values of $\Omega_m$ that are 
preferred in the presence of $\alpha_s$ than in the case of constant $n_s$. 
Without the SN data to restrict the matter density, small-scale structure suppression can be offset by an increase in the matter density. 
Figure (\ref{test4}) shows the absence of a degeneracy between $\alpha_s$ and $\sum m_{\nu}$. 
 
\subsection{Primordial tensor perturbations?} 
Temperature anisotropies are sourced not
only by density fluctuations (scalar modes) but by tensor modes
on small angular scales. Inflationary models predict their production by gravitational waves
and evolution independently of scalar modes, leading to an uncorrelated power spectrum 
\cite{souradeep}. On scales of the Hubble radius such modes interfere with the photon 
propagation along the line of sight and in so doing, induce extra anisotropy predominately on large angular scales.  
In the CMB, the presence of tensor
perturbations is felt on small angular scales with increased 
low-$\ell$ power as well as a lower amplitude of
density fluctuations $A_s$. 
Since $A_s$ affects the height of the first acoustic peak
we can use the LSS data to normalize the CMB power spectrum thereby finding the best fit to
the peak and the large scale part and in so doing can constrain the amplitude of the tensor 
contribution, $A_t$, and $A_s$ simultaneously. Different models predict varying amounts of tensor 
perturbations and hence it is parameterized in terms of its ratio to
the scalar mode anisotropies.  Given the degeneracy between
$A_t$ of these metric perturbations  with $\omega_m$, $\omega_b$ and $n_s$ \cite{Wang} 
it is important to check whether a higher contribution $r$ is tolerated in the 
presence of massive neutrinos.
We use the same convention as given in \cite{Leach} where $r$ is the 
ratio of the primordial amplitudes of the tensor and scalar fluctuations
\be
r\equiv \frac{P_{tens}(k_{*})}{P_{scalar}(k_{*})}
\ee Here $k_{*} = 0.002$ Mpc$^{-1}$. 
This parameter relates to the spectral index of the tensor modes in the following way;
\be
n_t = -r / 8 .
\ee
We do not allow for the running of the spectral indices which would reduce the amount of tensors that can fit the data.  
The upper limit on the neutrino mass degrades slightly to $\sum m_{\nu} = 1.38$ eV with the introduction of tensors with their 
fraction characterized by $r = 0.10^{+0.12}_{-0.08}$. Bearing in mind that the current theoretical models which support the presence 
of tensors are few, it is reasonable to conclude from this result that excluding tensor perturbations 
from the cosmological model will not effect the robustness of neutrino mass constraints.

\subsection{Mixed Initial Conditions?}
Up to this point we have assumed that the initial conditions in the
universe are purely adiabatic. When this assumption is relaxed, 4 isocurvature regular modes are assumed to arise, namely; CDM (CI),
baryon (BI), neutrino density (NID) and neutrino velocity (NIV) (where the momentum densities
of the neutrino and photon-baryon fluids are assumed to cancel exactly \cite{BMT}).
It was proposed in \cite{Parkinson} that the CMB and baryon isocurvature modes can not
be differentiated observationally and confirmed by the identical (up to an unimportant
multiplicative constant) correlation matrix elements \cite{TRD1}.
We consider the simultaneous mixing of all distinct modes so there will be 10 components of the initial
power spectrum matrix;  
\be
<A_i(\bar{k}) A^*_j(\bar{k}')> = P_{ij}(k)\delta^3 (\bar{k} - \bar{k}' ),
\ee where ($i , j = 1, 2, 3, 4, 5$ ) label the modes and their amplitudes $A_i$.  
Assuming that the initial power spectrum of each mode is a smoothly varying function
of $k$ and given by  $P_{ij}(k) =  A_{ij} k^{n_s }$, where each correlation is characterized by an amplitude $A_{ij}$ and a single scalar spectral
index (the auto-correlated adiabatic mode is already characterized by the parameter $A_s$).  
Following \cite{bucher2} it holds that
\be
A_{ij} \propto z_{ij}
\ee where
\be
trace\bra{z z^{T}} = \sum_{i, j = 1}^{N} z^{2}_{ij} = 1
\ee
So to characterize the all distinct isocurvature modes we introduce 9 additional parameters, 
namely the co-efficients $z_{ij}$ which give the relative contributions of each spectra to 
the normalization.  This parameterization is advantageous because in our case where $n_{ad} = n_{iso}$, it is independent of the value of the pivot scale $k_{0}$.

\subsubsection{Adiabatic +  CDM isocurvature}
We initially consider the simplest extension of the adiabatic regime, which is a single 
isocurvature mode mixed with the adiabatic mode \cite{beltran}.  Including only the CI and AD modes 
we arrive at the same constraints as in the adiabatic case of $1.27$ eV.  We do not find any 
inter-dependence between the parameters characterizing the CI modes and the neutrino mass. 

\subsubsection{AD + CI + NID}
The neutrino isocurvature modes are generally difficult to motivate theoretically. Vanishing neutrino 
chemical potentials mean that there are the same numbers of neutrinos as anti-neutrinos produced in each generation.  
The implication is that spatial differences in the densities of neutrinos and photons will be naturally erased via the processes involving $\nu$ $\bar{\nu}$ 
annihiliation. The netrino modes can thus only be generated at temperatures as low as a few MeV where these processes have ceased. 
Avoiding these issues requires the introduction of non-zero chemical potentials.  We find that when the NID isocurvature mode is admitted, significantly 
higher neutrino masses with an upper limit $2.4$ eV at the 95$\%$ CL are permissible when compared to our adiabatic constraint of $1.3$ eV.

\subsubsection{AD + all isocurvature modes} 
To generate the neutrino velocity mode, a spatially varying relative velocity is needed between the neutrinos and photons 
but must be constructed initially such that the overall perturbation in total momentum density is zero \cite{BMT}.  
Such modes therefore require even more exotic generation mechanisms and are met with more skepticism.  Adding in the neutrino 
velocity mode and permitting all 4 distinct isocurvature modes to mix,
leads to a slightly lower neutrino mass constraint of $2.2$ eV. 
\begin{figure}[t]
  \begin{center}
\includegraphics[trim = 0mm 50mm 0mm 0mm, scale=0.7]{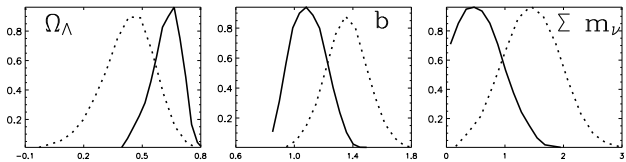}
  \end{center}
\caption{1D marginalized parameter distributions for adiabatic 
only (solid line) and a correlated mixture of adiabatic and
isocurvature initial conditions (dotted line). }
  \label{h_17d}
  \vspace{-.1in}
\end{figure}
The distribution of the galaxy bias as shown in figure (\ref{h_17d}) exhibits a shift to 
higher values than in the pure adiabatic case. Since 
the matter power spectrum is dominated by the adiabatic mode \cite{dunkley2}, 
including an isocurvature mode reduces the overall power $P(k)$, 
requiring a higher biasing of the galaxy spectrum to that of the underlying matter distribution. 
But because our value for $b$ agrees with the $1.3 \pm 0.2$ obtained for all 
4 correlated isocurvature modes in the presence of \emph{massless} neutrinos \cite{bucher2},
we infer that the bias is not sufficiently high to require suppression of the matter power by such high mass neutrinos.  
A lower value of $\Omega_{\Lambda}$ than that reported in \cite{bucher2}
must then compensate for the higher allowed value of neutrino mass. 
This is a strong indication of a degeneracy between $\sum m_{\nu}$ and a 
combination of isocurvature parameters. Note that the neutrino mass distribution in figure (\ref{h_17d}) 
seems to exclude the model with $\sum m_{\nu} = 0$. However this feature 
can be attributed to a larger prior space than the massless case in which 
NIV and NID modes are 0. In other words, there are effectively more ways of fitting 
the data with massive neutrinos.

\begin{table}
\begin{center}
\begin{tabular}{cc}    \hline
\textbf{Additional parameters}  &   $\sum m_{\nu}$ for $N_{\nu}=3.04$ \\
 & $95 \%$ CL 
\\ \hline
Spatial Curvature $\Omega_{K}$  &  $1.17$ eV\\
Dark energy $w$ ($w$ = constant)  &  $1.18$ eV\\
Tensors $A_t$ &    $1.38$ eV\\
Running spectral index $n_{run}$  &   $1.66$ eV\\
Isocurvature (all modes) &   $2.2$ eV \\
\hline
\end{tabular}
\end{center}
\caption{Constraints on neutrino mass for different extensions to parameter space for the fiducial CMB+LSS dataset.  }
\label{param_table}
\end{table}

\section{The path to tighter constraints}

If we consider the degredation of the neutrino 
mass constraints which results as the hypothesis space enlarges, it seems important to 
determine how we can improve our knowledge of $\sum m_{\nu}$ in the presence of these additional parameters. 
We have identified two high strong degeneracies between $\sum m_{\nu}$ and $\Omega_m$ and $b$ which become important when seeking stronger constraints. 
Any supplementary information that reduces the freedom in these degenerate parameters will
impact on the neutrino constraints. This information must come from independent experiments 
which measure phenomena on which the neutrino mass have negligible effect.  

Stringent neutrino mass constraints can be arrived at by limiting the allowable values of $b$. The upper bound of $\sum m_{\nu}< 3$ eV (95 $\%$ 
CL found by Hannestad \cite{Hannestad} using the WMAP 1st-year \cite{WMAP} and 2dFGRS \cite{2df} datasets leaving $b$ and 
the normalization as free parameters was lowered to $0.7$ eV by the WMAP team using the same data by assuming $b \simeq 1$, 
based on the bi-spectrum analysis of the 2dFGRS \cite{Verde_lahav}. Using most recent CMB+LSS data, we update this bound to $0.63$ eV.  
The effects of introducing different measurements of the bias are illustrated in table (\ref{bias_table}). We wish to dissect the 
effect of including bias measurements on neutrino mass constraints in order to establish how they need to be improved to better pin down  $\sum m_{\nu}$. 

\begin{figure}[t]
  \begin{center}
\includegraphics[trim = 0mm 15mm 0mm 0mm, scale=0.7]{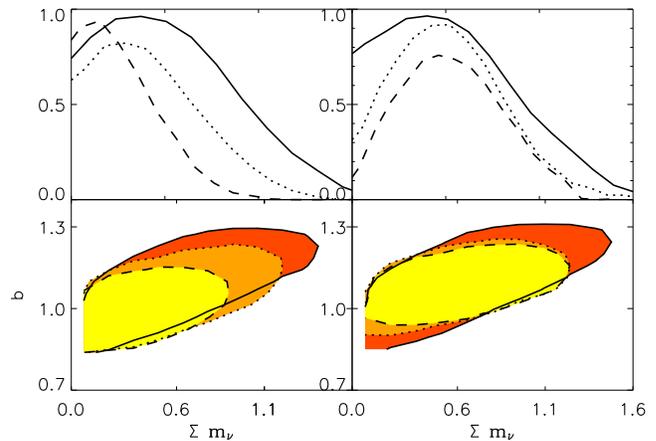}
  \end{center}
\caption{
 \emph{Bottom panel:} The 95 $\%$ constraints in the  ($\sum m_{\nu}$, $b$) plane for different Gaussian 
priors on the bias. On \emph{left}, $\mu_b$ is $1.1$ (solid), $1.0$ (dotted) and $0.92$ (dashed). On the \emph{right}, 
$\sigma_b$ is $0.11$ (solid), $0.0825$ (dotted) and $0.055$ (dashed). 
\emph{Top panel:} The 1D marginalized distribution for $\sum m_{\nu}$ for the same cases.}
  \label{3m}
   \vspace{-.2in}
\end{figure}

When the analysis is performed using the SDSS data its natural distribution is found to be normal with 
mean $b=1.09$ and width $\sigma_b =  0.11$.  If we impose different Gaussian priors on $b$ keeping the mean
fixed and reducing the standard deviation to $0.75 \sigma_b$ and $0.5 \sigma_b$ , the resulting 1D $\sum m_{\nu}$ distributions tighten slightly.  
Because the peaks shift towards higher values as shown in figure (\ref{3m}) the upper bounds at $95\%$ CL are not lowered one might expect. 
With reference to allowed regions in ($\sum m_{\nu}$, $b$) space, the effect of the imposed standard deviation of $b$ on the mass 
limit progressively weakens, indicative of a threshold value of $\sigma_b$ beyond which it has no effect at all. If we now lower the mean values $\mu_b$ 
of the priors but where $\sigma_b= 0.1 \times\mu_b$, the allowable region in ($\sum m_{\nu}$, $b$) space is reduced in a relatively linear fashion.  
Given the correlation between $b$ and $\sum m_{\nu}$, a 
decrease in $\mu_b$ means that a lower $b$ is also favoured and is equivalent to weighting the neutrino mass around a correspondingly lower value. 
This shows that in order to improve our knowledge of $\sum m_{\nu}$ a more \emph{precise} measurement of the mean $b$ 
from future analyses such as weak lensing, will be helpful. 

In light of recent results \cite{Percival} the assumption that power spectrum of the 
galaxy distribution is \emph{linearly} biased relative to the matter power spectrum may be short-sighted, with 
implications for current neutrino constraints. The lower matter density favoured by the SDSS DR5 galaxy data could be explained in terms of a scale-dependent bias. 
This has the potential to degrade neutrino constraints because 
the effect of the bias will no longer be confined to the amplitude but will now be felt by the \emph{shape} 
of the linear power spectrum, making it harder to break the degeneracy with $\sum m_{\nu}$.
In work \cite{Kristiansen} in which the authors assess the effect of different datasets on the neutrino mass limits,  
discrepancies were pointed out between upper bounds on $\sum m_{\nu}$ obtained using the CMB data and either the LSS measurements from 2DF \cite{2dF_05} or SDSS \cite{SDSS}. 
Normalizing the \emph{total} matter power spectrum without having to make such assumptions can be achieved from weak lensing experiments. 
The limit on the neutrino mass found using the measurement of the cluster mass function in \cite{dahle} and the CMB data was considerably weaker. 

\begin{table}[t]
\begin{center}
\begin{tabular}{cc}    \hline
\textbf{Dataset}  &   $\sum m_{\nu}$ for $N_{\nu}=3.04$ \\
 & $95 \%$ Confidence Level (CL) 
\\ \hline
CMB  &  $2.0$ eV\\
CMB +LSS &    $1.3$ eV\\
CMB+LSS+2dF bias \cite{Verde-et al} &   $1.05$ eV\\
CMB+LSS+SDSS bias \cite{seljak} &   $0.79$ eV\\
CMB+LSS+ $b=1$ &   $0.63$ eV\\
\hline
\end{tabular}
\end{center}
\caption{Constraints on neutrino mass for different information about the bias $b$ for the 
standard 8-dimensional parameter space. }
\label{bias_table}
\end{table}

\section{Conclusions}
In this paper, the dependance of constrains on the neutrino mass using cosmological observations (CMB+LSS) on the underlying model has been assessed. 
Before proceeding with the analysis we systematically reviewed most decisions that go into such an investigation and are satisfied that analysis with 
current datasets is unaffected when different neutrino parameterizations are employed. 
The BAO \cite{BAO} and SN \cite{astier} results are shown to impact the mass limits signficantly and are thus incorporated with caution.

When parameter space is extended beyond our current standard cosmological model ($\Lambda$CDM), we rule out any degeneracy between neutrino mass and spatial curvature. 
We also find that the relationship between dark energy equation of state $w$ and $\sum m_{\nu}$ 
depends on the datasets used. A decrease in $w$ away from $-1$ can be accommodated by an increase in the neutrino mass or 
an increase in $\Omega_m$ which rapidly becomes incompatible with the SN data.   The inclusion of a running spectral 
index no longer tightens the upper bound on $\sum m_{\nu}$ if the SN measurements are removed from the dataset. Both of
these mentioned results disagree with former findings in analyses which include the SN data and highlight issues regarding the 
dependence of the results not only on the cosmological model but the cosmological observations used  (see \cite{Kristiansen}). 
Allowing for tensor perturbations is found to have minimal effect, however, the analysis excluded the possibility of a running spectral index.
The substantial degradation of the upper mass bound to $2.2$ eV with more general initial conditions points to a degeneracy with the isocurvature parameters, 
namely the neutrino density mode.  These high-$\sum m_{\nu}$ models are however ruled out by the SN constraints 
on $\Omega_m$ \cite{astier} which bring the limit down to $1.17$eV.    Constraining the value of bias impacts significantly on the upper bound, with the most stringent constraint 
of $\sum m_{\nu} < 0.63$ eV being obtained when taking $b=1$. Assuming a known distribution for the bias amounts to effectively incorporating a data set and all accompanying 
systematic uncertainties. For this reason the same discernment should be exersized.  We find that improvements in the accuracy 
of the measurements of the galaxy bias beyond a point will not dramatically aid our constraints on $\sum m_{\nu}$. 
In conclusion it has been shown that the cosmological constraint on 
neutrino mass is sensitive to many factors and it is only once 
\emph{all} assumptions have been evaluated can we regard the resulting limit as robust.

{\it Acknowledgments}:
We are extremely grateful to J. Dunkley
for guidance in this project. We thank S. Biller and K. Moodley for
discussions. CZ acknowledges support from a Domus A scholarship awarded by Merton College.

\end{document}